\documentclass[aps,twocolumn,floatfix,altaffilletter,superscriptaddress,preprintnumbers,tightenlines,showpacs,showkeys,notitlepage,nofootinbib]{revtex4-1}

\usepackage[T1]{fontenc}
\usepackage[colorlinks=true,citecolor=blue,linkcolor=blue]{hyperref}
\usepackage{mathrsfs}
\usepackage{braket,graphicx,physics,amsthm,amssymb,amsfonts,xcolor,booktabs,siunitx,caption,subcaption}
\usepackage{mathtools}
\usepackage{slashed}
\usepackage[capitalise, english]{cleveref}
\usepackage[normalem]{ulem}
\usepackage{slashed}	
\usepackage{xspace}
\newcommand{\muthreee}{\textit{Mu3e}\xspace}

\definecolor{cborange}{HTML}{e69f00}
\definecolor{cbgreen}{HTML}{009e73}
\definecolor{cbyellow}{HTML}{f1dd42}
\definecolor{cblblue}{HTML}{56b4e9}
\definecolor{cbblue}{HTML}{0072b2}
\definecolor{defgrey}{HTML}{9f9f9f}
\definecolor{defgreen}{HTML}{8eba42}

\begin{document}

\preprint{}

\title{
A Robust Search for Lepton Flavour Violating Axions at Mu3e
}

\author{Simon Knapen}
\affiliation{Berkeley Center for Theoretical Physics, University of California, Berkeley, CA 94720, U.S.A.}
\affiliation{Theory Group, Lawrence Berkeley National Laboratory, Berkeley, CA 94720, U.S.A.}

\author{Kevin Langhoff}
\affiliation{Berkeley Center for Theoretical Physics, University of California, Berkeley, CA 94720, U.S.A.}
\affiliation{Theory Group, Lawrence Berkeley National Laboratory, Berkeley, CA 94720, U.S.A.}

\author{Toby Opferkuch}
\affiliation{Berkeley Center for Theoretical Physics, University of California, Berkeley, CA 94720, U.S.A.}
\affiliation{Theory Group, Lawrence Berkeley National Laboratory, Berkeley, CA 94720, U.S.A.}

\author{Diego Redigolo}
\affiliation{INFN, Sezione di Firenze Via G. Sansone 1, 50019 Sesto Fiorentino, Italy}

\begin{abstract}
We propose a search at \muthreee for lepton flavor violating axion(-like) particles in $\mu\to 3e + a$ decays. By requiring an additional $e^+e^-$ pair from internal conversion, one can circumvent the calibration challenges which plague the $\mu \to e+a$ channel for axions lighter than $\SI{20}{\MeV}$. Crucially, the corresponding reduction in signal rate is to a large extent compensated for by \muthreee's ability to resolve highly collimated tracks. For phase I of \muthreee, we project a sensitivity to decay constants as high as $\SI{6E+9}{\GeV}$ which probes uncharted parameter space in scenarios of axion dark matter. The sensitivity to axions which couple primarily to right-handed leptons can be further improved by leveraging the polarisation of the muon beam.  
\end{abstract}


\maketitle

\section{Introduction}
The Standard Model (SM) features a large $SU(3)^5$ flavour symmetry, broken (albeit weakly) by the Yukawa couplings. The precise origin of this breaking remains one of the main unsolved mysteries in particle physics. Should this breaking be spontaneous in nature, one would expect a collection of extremely weakly coupled pseudo-Nambu-Goldstone bosons (pNGb), many of which may well be light~\cite{Wilczek:1982rv,Reiss:1982sq, Gelmini:1982zz, Feng:1997tn}. Concrete realizations of this general class of pNGb's  are familons in Frogatt-Nielsen models of flavor~\cite{Froggatt:1978nt}, as well as majorons in models addressing the origin of the neutrino masses~\cite{Calibbi:2020jvd, Ibarra:2011xn, Garcia-Cely:2017oco, Heeck:2019guh}. Moreover, any of these pNGb's may play the role of the QCD axion, if the Peccei-Quinn symmetry has a non-trivial embedding in the SM flavour group~\cite{Ema:2016ops, Calibbi:2016hwq, Linster:2018avp, Calibbi:2020jvd}. 

In this letter, we consider the possibility of such an axion-like particle, denoted as $a$, with flavour-violating couplings to the muon and electron of the form
\begin{align}\label{eq:model}
\mathscr{L}_\text{FV-a} &= \frac{m_\mu}{2 f_a} \frac{1}{|C_{e\mu}|} a \bar \mu (C^V_{e\mu}+ C^A_{e\mu} \gamma_5) e + \text{h.c.}\,,
\end{align}
with \mbox{$|C_{e\mu}|\equiv \sqrt{(C^V_{e\mu})^2+(C^A_{e\mu})^2}$}, where one or both $C^{V,A}_{e\mu}$ are non-zero. The scale $f_a$ is the decay constant associated with the breaking of the flavour symmetry in the ultraviolet. The coefficients $C^{V,A}_{e\mu}$ are model-dependent and parametrize the low energy imprint of the full theory. If the mass of the axion-like particle, $m_a$, is less than $m_\mu - m_e$, Eq.~\eqref{eq:model} induces exotic muon decays of the form $\mu\to e+a$. For a given limit on this exotic branching fraction, the extraordinarily small width of the muon already affords exclusion limits at large values of \mbox{$f_a\gtrsim 10^9$ GeV} for $C^{V,A}_{e\mu}\sim 1$. 

The axion's lifetime on the other hand is determined by the size of its flavor diagonal couplings, either to photons or electrons. If these couplings are assumed to be as small as the flavor violating ones, an axion lighter than \SI{1}{\MeV} would easily live for several hours or more before decaying. It is therefore expected to be stable and invisible at typical detector length scales, unless a highly unnatural hierarchy is engineered between the flavor-diagonal and flavor-violating couplings. While such a hierarchy can be constructed, such a model is expected to generate tension with other searches for charged lepton flavor violation, such as $\mu^+\to e^+\gamma$, $\mu^+\to e^+ e^- e^+$ and $\mu\to e$ conversions. Finally, the axion $a$ could also be stable on the scale of the Universe today, in which case it could play the role of the dark matter either produced through a combination of thermal \cite{DEramo:2021usm,Panci:2022wlc} and non-thermal mechanisms~\cite{Calibbi:2020jvd}.

For the above reasons, searches for invisible axions in $\mu \to e+a$ decays were carried out at TRIUMF \cite{Jodidio:1986mz, TWIST:2014ymv}, yielding limits as high as $f_a\gtrsim \SI{2.45E+9}{\GeV}$. 
\begin{figure}
    \centering
    \includegraphics[scale = 1]{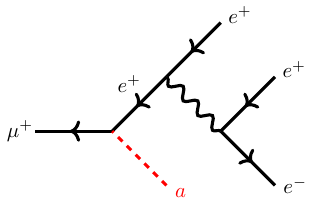}
    \caption{Representative diagram for the $\mu\to3e+ a$ decay, with an additional $e^+e^-$ pair from the internal conversion of a virtual photon. We presume $a$ to be invisible.}
    \label{fig:diagram}
\end{figure}
The upcoming \muthreee experiment at the Paul Scherrer Institute (PSI) is ideally suited to carry this program forward with a nearly hermetic tracker and expecting well over $10^{15}$ muons on target.
The strategy is to search for a monochromatic line on top of the Michel spectrum \cite{Perrevoort:2018ttp, Perrevoort:2018okj}. This method is particularly powerful for $m_a\gtrsim \SI{25}{\MeV}$, but runs into complications at lower masses. This is primarily because the monochromatic line now lies very close to the kinematic edge of the Michel spectrum, which serves as the basis for calibrating the detector. Consequently, an alternative calibration technique and/or search strategy is required.
\begin{figure*}
    \begin{center}
        \includegraphics[width = 0.495\textwidth,trim={0.45cm 0 0.4cm 0}, clip]{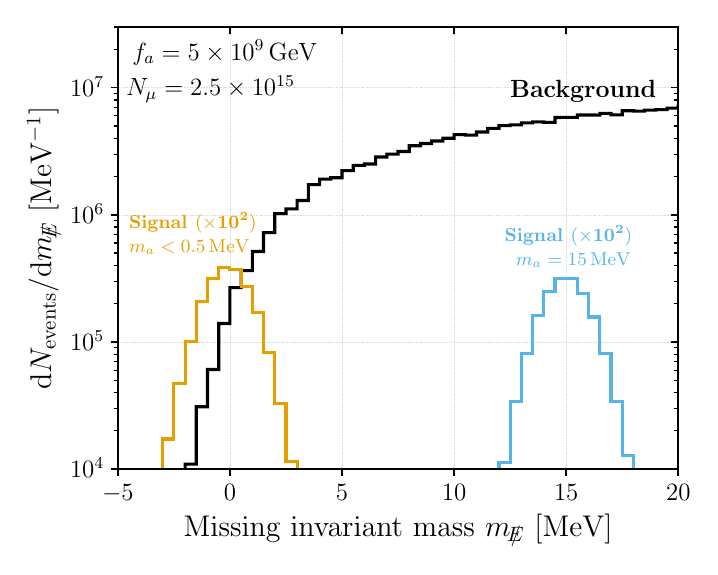}\hfill
        \includegraphics[width = 0.495\textwidth,trim={0.45cm 0 0.4cm 0}, clip]{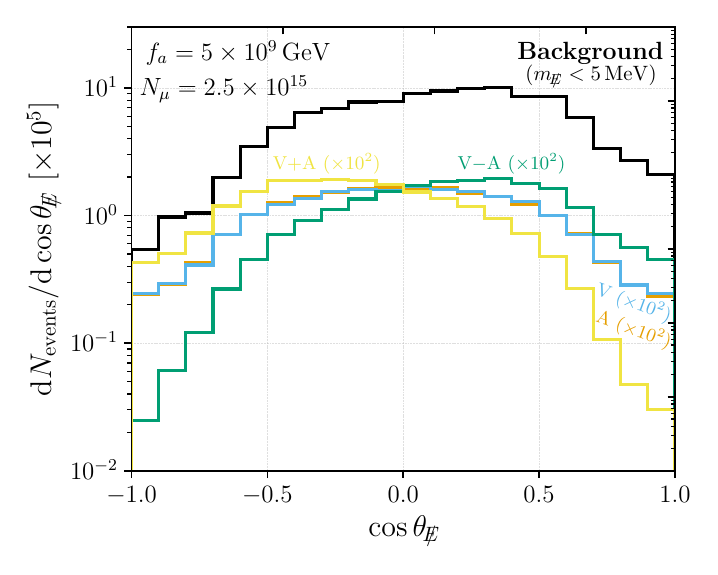}
    \end{center}
    \caption{\textbf{Left:} The missing mass, denoted as $m_{\slashed{E}}$, distribution is shown for two signal benchmark points: a light axion with a mass below the detector resolution in {\color{cborange}\bf orange} and a $\SI{15}{\MeV}$ axion in {\color{cbblue}\bf blue}. The $\mu\to3e+2\nu$ background is shown in {\bf black}. It's important to note that due to the finite track momenta resolution, $m_{\slashed{E}}$ can be reconstructed as a negative number (see \cref{foot:res}).
    \textbf{Right:} This panel displays the angle of the missing momentum vector with respect to the beam line and polarisation vector ($\cos \theta_{\slashed{E}}$) for four signal chirality structures with $m_a=0$. Here, the $A$ structure is represented in {\color{cborange}\bf orange}, the $V$ structure in  {\color{cbblue}\bf blue}, the $V-A$ structure in {\color{cbgreen}\bf green}, and the $V+A$ structure in {\color{cbyellow}\bf  yellow}. The background distribution is limited to events with $m_{\slashed{E}}<\SI{5}{\MeV}$. Both panels are based on the phase~I of \muthreee with $N_\mu = \SI{2.5E+15}{}$ muons on target.}
    \label{fig:costheta_dist}
\end{figure*}

In this Letter we propose an alternative search strategy which circumvents the need for a new calibration technique, by using the sub-leading $\mu \to 3e +a$ channel, where an additional $e^+e^-$ pair originates from the internal conversion of a virtual photon (\cref{fig:diagram}). As a result none of the three electron momenta are close to the kinematic edge of the Michel spectrum. This trades a potentially irreducible systematic limitation for an effective reduction in statistics. The latter is however ameliorated by the exquisite sensitivity of \muthreee to collimated tracks, as both signal and background are enhanced in the colinear part of the phase space.

\begin{figure*}
    \centering
    \includegraphics[width=0.8\textwidth]{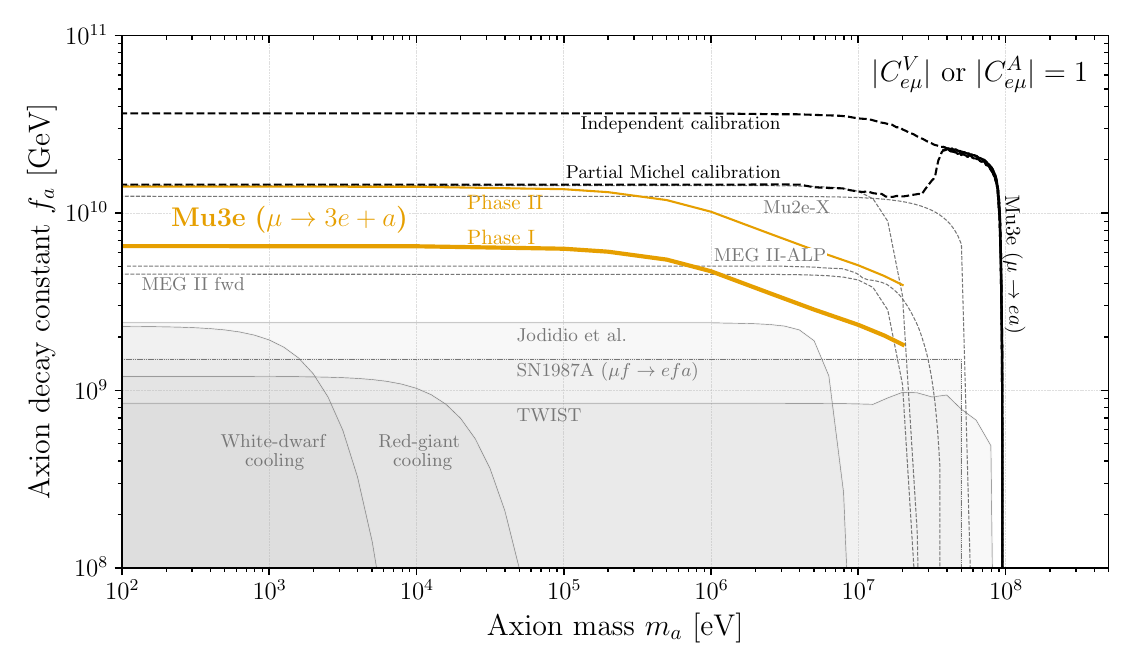}
    \caption{Projected sensitivity of the $\mu\to3e+a$ channel at \muthreee ({\color{cborange}\bf orange lines}) for a flavour-violating axion-like particle with axial couplings ($C^A_{\mu e}=1, C^V_{\mu e}=0$), along with existing bounds and other projected sensitivities. The $\mu\to3e+a$ channel can still be used to search for axions with $m_a>\SI{20}{\MeV}$, but in this region the normal $\mu\to e+a$ search should always be superior and we therefore terminate the yellow line. The phase I (phase II) limits correspond to $N_\mu = \num{2.5E+15}$ ($\num{5.5E+16}$) muons on target. We display existing bounds ({\color{defgrey}\bf grey shaded regions}) from muon factories, Jodidio et al.~\cite{Jodidio:1986mz} and TWIST \cite{TWIST:2014ymv}, the supernova SN1987a bounds ($\mu f \to e f a$) \cite{Zhang:2023vva}, bounds from the anomalous cooling of white dwarfs \cite{MillerBertolami:2014rka}, and red giants \cite{Raffelt:1994ry,Viaux:2013lha}. Two {\bf dashed-black lines} represent the projected sensitivity for $\mu\to e+a$ at \muthreee, reflecting a hypothetical detector calibration independent of the Michel spectrum endpoint and a second calibration relying purely on the low-energy part of the Michel spectrum \cite{Perrevoort:2018ttp,Perrevoort:2018okj}. The \emph{Mu2e}-X curve shows the estimated sensitivity on $\mu\to e+a$ of a calibration sample of $\num{3E+13}$ stopped muons~\cite{Hill:2023dym}.The \emph{Mu2e}-X reflects a projection that is slightly updated relative to the result in~\cite{Hill:2023dym}. These reaches do not take into account experimental systematics which are likely to limit the reach at low $m_a$ (see text for additional details). The \emph{MEG II-ALP} curve reflects the estimated sensitivity of a dedicated, low luminosity run (for a duration of 1 month) with the standard \emph{MEG II} detector configuration. The \emph{MEG II-fwd} curve refers to a proposal to install a dedicated forward detector at \emph{MEG II} whose sensitivity strongly depends on the details of the configuration~\cite{Calibbi:2020jvd}.
    }
    \label{fig:final-constraints}
\end{figure*}

In the remainder of this letter, we investigate the sensitivity of this strategy compared with the projected sensitivity of the $\mu \to a+e$ channel either at \muthreee~\cite{Perrevoort:2018ttp, Perrevoort:2018okj} or at \emph{Mu2e}~\cite{Hill:2023dym}, as well as a recent proposal to search for $\mu \to a+e+\gamma$ at \emph{MEG II} during a dedicated, low-luminosity run \cite{Jho:2022snj}. Our results are summarised in Fig.~\ref{fig:final-constraints}.

\section{Analysis strategy}
We used the \texttt{FeynRules 2.3} package \cite{Alloul:2013bka} to generate a \texttt{Universal FeynRules Output} \cite{Degrande:2011ua} model file for the model defined by \cref{eq:model}. We then used \texttt{MadGraph5\_aMC@NLO 3.4.1} \cite{Alwall:2014hca} to generate the signal events, accounting for the polarisation of the muon. Similarly, the $\mu \to 3e + 2\nu$ background was generated with \texttt{MadGraph5\_aMC@NLO 3.4.1}, again accounting for the muon polarisation. We assumed that other backgrounds, such as Bhabha scattering, external conversions and random coincidences, are negligible compared to $\mu \to 3e + 2\nu$ after applying vertex quality cuts, as is the case in the standard $\mu\to3e$ analysis \cite{Mu3e:2020gyw} and searches for dark photons \cite{Perrevoort:2018okj,Echenard:2014lma}. As a benchmark, we assumed that the muons are 86\% polarised, as was measured at MEG~\cite{MEG:2015kvn}. However, this value is experiment-dependent and will need to be measured by the collaboration. We assumed a sample of $2.5\times 10^{15}$ ($5.5\times 10^{16}$) muons stopped in the target for phase I (phase II) of \muthreee~\cite{Mu3e:2020gyw}. For phase II, there is also expected to be a significant improvement in the detector itself, though the number of coincident Michel decays will also increase. We do not include these changes in our analysis as they have not yet been finalised by the \muthreee collaboration. Instead we simply extrapolate the phase I sensitivity to the larger number of muons that will be collected during phase II.

Our modelling of the \muthreee detector acceptance and resolution is explained in detail in our companion paper, Ref.~\cite{Mu3e-angles}. For this analysis, the most important features are the following: ($i$) tracks with transverse momentum less than \SI{12}{\MeV} were assumed not to be reconstructed, ($ii$) the track and vertex efficiencies were taken to be the same as those projected for the $\mu\to3e$ analysis \cite{Mu3e:2020gyw}, and ($iii$) the resolution on the missing mass was modelled as Gaussian-distributed with a mean ranging between \SI{1.21}{\MeV} and \SI{0.9}{\MeV}, depending predominantly on the number of short/long tracks reconstructed. To estimate the missing mass resolution, we used the resolution on $m_{ee}$, the invariant mass of an electron-positron pair, from the $\mu \to e \bar{\nu} \nu X (X\to e^+e^-)$ search in Fig.~(6.12a) of Ref.~\cite{Perrevoort:2018okj} as a proxy. This resolution is almost flat as a function of $m_{ee}$, allowing extrapolation to the low missing mass required here. Note that without information on the angular resolution as a function of energy, a more accurate determination of the missing mass resolution is not possible.
\begin{figure*}
    \centering
    \includegraphics[width=0.495\textwidth,trim={0.4cm 0 0.0cm 0}, clip]
    {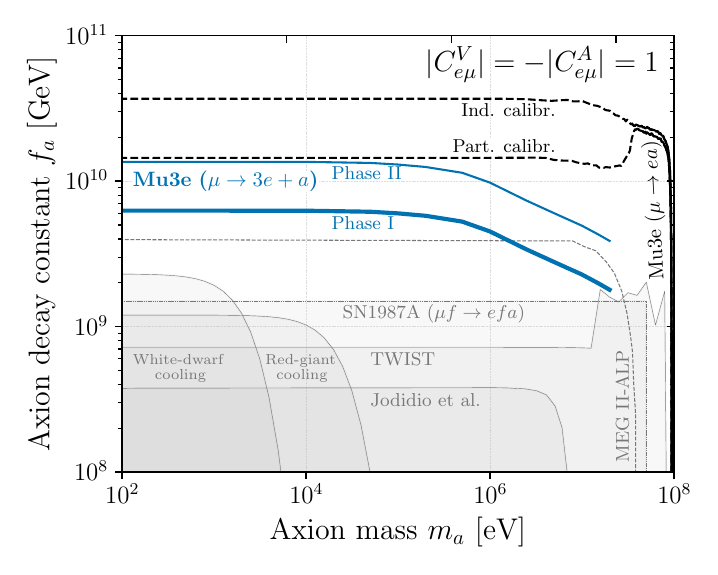}
    \hspace{-0.3cm}
    \includegraphics[width=0.495\textwidth,trim={0.4cm 0 0.0cm 0}, clip]{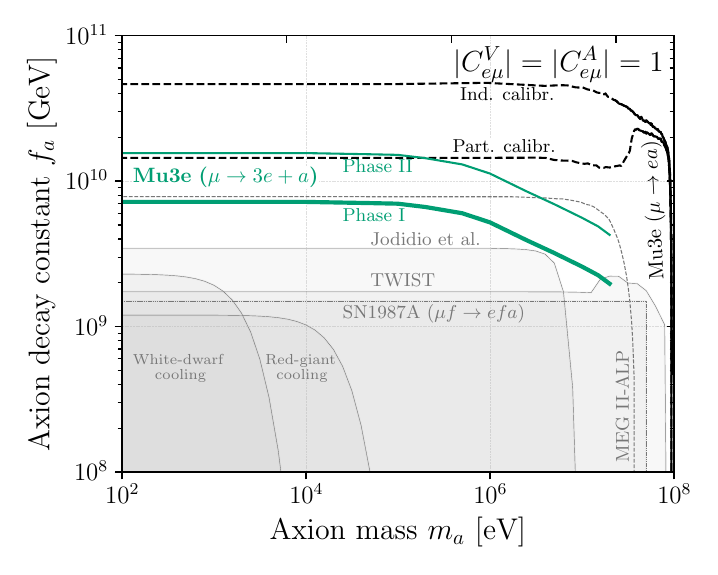}
    \caption{Projected sensitivities of the $\mu\to3e+a$ channel at \muthreee for a flavour-violating axion-like particle with left ($V-A$) and right ($V+A$) chiral structure in {\color{cbblue}\bf blue} and in {\color{cbgreen}\bf green} respectively, along with existing bounds ({\color{defgrey}\bf grey shaded})  and other projected sensitivities. Note that the projections for \muthreee in the $\mu \to e + a$ channel using part of the Michel spectrum were derived only for $C_{e\mu}^A = 1$ \cite{Perrevoort:2018ttp,Perrevoort:2018okj}. For additional details on the other searches see \cref{fig:final-constraints}.}
    \label{fig:final-constraints-chiral-opts}
\end{figure*}
By assuming that the muon decayed at rest, one can reconstruct\footnote{Due to the finite resolution of track momenta, the square of the invariant mass $m_{\slashed{E}}^2$ can be reconstructed as a negative number. To allow for this possibility, we defined $m_{\slashed{E}}\equiv \text{sign}(m_{\slashed{E}}^2)\sqrt{|m_{\slashed{E}}^2|}$, hence $m_{\slashed{E}}$ can take negative values, as seen in \cref{fig:costheta_dist}.\label{foot:res}} the missing mass $m_{\slashed{E}}$, which at truth-level corresponds to $m_a$ for the signal. The $m_{\slashed{E}}\approx 0$ regime is also the kinematic endpoint for the background, where both neutrinos are collinear. This implies that the background drops quickly as $m_{\slashed{E}}\to 0$, affording excellent sensitivity, as is evident from the left-hand panel of \cref{fig:costheta_dist}.

Secondly, we consider the angle between the missing momentum vector and the beamline ($\cos \theta_{\slashed{E}}$), as shown in the right-hand panel of \cref{fig:costheta_dist}. In this panel, a slight tilt in the background distribution is observed, which can be attributed to the left-handedness of the weak interaction and the polarisation of the muons. Similarly, the signal distributions exhibit a tilt for $V-A$ ($C^V_{e\mu}=-C^A_{e\mu}$) and $V+A$ ($C^V_{e\mu}=C^A_{e\mu}$) couplings, while the vector $V$ ($C^V_{e\mu}=1, C^A_{e\mu}=0$) and axial $A$ ($C^V_{e\mu}=0, C^A_{e\mu}=1$) cases are symmetric in $\cos \theta_{\slashed{E}}$. Other kinematical variables, such as the magnitudes and angles of the various tracks and missing momenta, were systematically explored, as well as pairwise invariant masses and angles between each track and the missing momentum. All of these variables were found to be strongly correlated with $m_{\slashed{E}}$ and $\cos \theta_{\slashed{E}}$, and were therefore not included in the analysis for simplicity. It is possible that the inclusion of the full kinematics of the events could lead to slightly better sensitivity than what is projected here, although we anticipate any additional gain to be minor.

To estimate the sensitivity, we construct a two-dimensional, binned Poissonian likelihood ratio in $m_{\slashed{E}}$ and $\cos \theta_{\slashed{E}}$. From this likelihood ratio, we extract an exclusion limit at the 95\% confidence level, where the number of observed events in the likelihood ratio is randomly drawn from the background-only distribution. This procedure is repeated $10^4$ times, and we take the median of the resulting distribution of limits as our expected exclusion limit. We have further assumed that the search would be limited by statistical rather than systematic uncertainties, which should be a good assumption for the final state considered.
\section{Discussion and results}
Our sensitivity estimate for $\mu \to 3e + a$ at \muthreee is shown in \cref{fig:final-constraints} for the axial coupling ($C^A_{\mu e}=1,C^V_{\mu e}=0$), along with the existing bounds and other available projections. The results for the remaining chirality structures are shown in \cref{fig:final-constraints-chiral-opts}. For $m_a < \SI{1}{\MeV}$ we project the expected  sensitivity to \mbox{$f_a \lesssim \SI{6.4E+9}{\GeV}$}. The sensitivity for right-handed axion-like particles ($V+A$) is slightly better than for the remaining chirality structures. This is due to the asymmetry shown in the right-hand panel of \cref{fig:costheta_dist} and is therefore dependent on the degree to which the muons will remain polarised in the experiment. These bounds are powerful enough to robustly exclude $a$ as a dark matter candidate if it is produced through freeze-in \cite{Panci:2022wlc}. If $a$ is instead produced through oscillations during the radiation dominated epoch, \muthreee will be sensitive to part of the parameter space between~$\SI{3}{\eV}\lesssim m_a \lesssim \SI{1}{\keV}$~\cite{Calibbi:2020jvd}.

For the $\mu\to e + a$ channel, also at \muthreee, we show two separate projections for phase I, taken from \cite{Perrevoort:2018ttp,Perrevoort:2018okj}. Both outperform our method at face value: Since both signal and background rates are suppressed for $\mu \to 3e + a$ relative to $\mu \to e + a$, the latter is always expected to be statistically more powerful. On the other hand, a $\mu \to 3e + a$ analysis can be performed offline, while the $\mu \to e + a$ search must be conducted in real time. This means that an improved event reconstruction is likely possible.\footnote{We thank Ann-Kathrin Perrevoort for pointing this out to us.} 

Moreover for low $m_a$, there are additional subtleties associated with the $\mu \to e + a$ channel which cannot be quantified easily prior to the commissioning of the detector. In particular, for $m_a\lesssim \SI{25}{\MeV}$ the positron energy will lie close to the Michel edge, which will nominally be used to calibrate the detector. The ``independent calibration'' line in \cref{fig:final-constraints} assumes that an alternative, independent calibration process can be found, such as Mott or Bhabha scattering (see for example Ref.~\cite{Rutar:2016ozg}). The ``Partial Michel calibration'' line assumes that the detector calibration can be performed on the Michel spectrum, excluding the region near the edge. In this case, the simulation of the detector response must be used to extrapolate to the endpoint of the steeply falling distribution, potentially introducing additional systematic uncertainties. Both projections currently only include statistical uncertainties and it is therefore conceivable that their true reach will be substantially weaker once the systematic uncertainties can be quantified. Should this prove to be the case, the $\mu \to 3e + a$ channel will provide a useful alternative which is less sensitive to systematic uncertainties.

Finally, we comment on similar searches proposed at other experiments. Firstly, the calibration data of \emph{Mu2e} can be used to hunt for $\mu^+\to e^+ + a$ and $\pi^+\to e^+ + a$, as suggested in Ref.~\cite{Shihua:thesis,Hill:2023dym}. An analogous search was proposed for the \emph{COMET} experiment but with $\mu^-$ decaying in orbit \cite{Xing:2022rob}. The acceptance will be very small for \mbox{phase I} of the experiment, though \emph{COMET} could be competitive during its phase II. Assuming one can avoid placing a blocker at the exit of their solenoid, they project a limit similar to the \muthreee projection with an independent calibration.
For low $m_a$, both \emph{Mu2e} and \emph{COMET} proposals are likely to encounter similar systematic uncertainties as the $\mu^+\to e^++a$ search at \muthreee.

The \emph{MEG II} detector can look for $\mu \to  e + \gamma + a$ if it performs a dedicated run with lower beam intensity \cite{Jho:2022snj}. The estimated reach shown in \cref{fig:final-constraints} assumes that such a special run could integrate data for one month. This may be too optimistic given the competition of this dedicated run with the \emph{MEG II}'s flagship measurement of $\mu^+\to e^+\gamma$ and should be re-scaled appropriately once the dedicated \emph{MEG II} data-set will have been collected. This will likely leave interesting parameter space to be probed by the $\mu \to 3e + a$ channel at \muthreee. 

Lastly, a more ambitious possibility would be to add a dedicated calorimeter in the MEG~forward region and employ a strategy similar to the old Jodidio et al.~measurement~\cite{Jodidio:1986mz}. The success of such a strategy crucially depends on the efficiency in selecting a 100\% polarised muon sample and on the momentum resolution of the forward calorimeter and should be assessed once a concrete experimental proposal will be put together. We show in Fig.~\ref{fig:final-constraints} the estimated sensitivity in Ref.~\cite{Calibbi:2020jvd} assuming a calorimeter with a 10 cm diameter and a percent level momentum resolution will be installed in the forward region of MEG~II and a purely polarized muon sample up to percent level depolarization effects with $10^{14}\, \mu^+$ will be achieved. As one can see the $\mu \to 3e + a$ channel at \muthreee outperforms this proposal for most of the ALP masses without introducing the same experimental challenges. 

\section*{Acknowledgements}
We thank Ann-Kathrin Perrevoort for feedback on \muthreee as well as both Jure Zupan and Ann-Kathrin Perrevoort for comments on the manuscript. We also thank Jure Zupan for providing us the updated \emph{Mu2e}-X projection. Lastly, we thank Dean Robinson, Zoltan Ligeti, Jure Zupan, Ann-Kathrin Perrevoort and Yongsoo Jho for useful discussions. The work of SK was supported by the Office of High Energy Physics of the U.S. Department of Energy under contract DE-AC02-05CH11231. KL was in part supported by the Berkeley Center for Theoretical Physics. TO is supported by the DOE Early Career Grant
DESC0019225. Part of this work was performed at the Aspen Center for Physics, which is supported by National Science Foundation grant PHY-1607611.
\appendix
\section{Additional analysis details}
\label{app:addition_analysis}

\begin{figure}
\includegraphics[width=0.495\textwidth,trim={0.45cm 0 0.3cm 0}, clip]{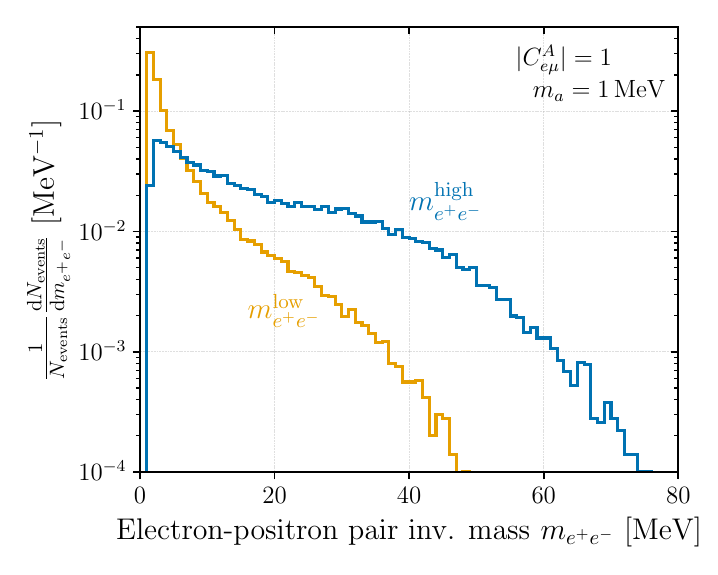}
\caption{Truth-level low ({\color{cborange}\bf orange}) and high ({\color{cbblue}\bf blue}) invariant mass combinations for both $e^+e^-$ pairs, for a signal benchmark with $m_a=\SI{1}{\MeV}$. Both distributions are peaked close to zero, which is a manifestation of the soft and colinear enhancements in the decay amplitude.}
\label{fig:mee}
\end{figure}
As explained in the main text, we focus exclusively on $m_a<\SI{20}{\MeV}$, as the regular $\mu \to e+a$ search should always be more sensitive for larger $m_a$. The truth-level $\mu \to 3e+a$ branching ratio is given by
\begin{align}\label{eq:brratio}
\text{Br}_{\mu \to 3e+a} \approx 2.5\times 10^{-10}\times\left(\frac{5\times 10^9\,\text{GeV}}{f_a}\right)^2\,,
\end{align}
for $m_a=\SI{1}{\MeV}$, where we required each track to have a transverse momentum exceeding \SI{10}{\MeV} ($p_T>\SI{10}{\MeV}$). For the parameter space of interest, the dependence of this branching ratio on $m_a$ is very mild. E.g.~for $m_a=\SI{20}{\MeV}$, the branching ratio is roughly 30\% lower than what is shown in \eqref{eq:brratio}. The detector efficiency is $30\%$ across all signal points we consider. It is dominated by the geometric acceptance for the tracks and their reconstruction efficiency \cite{Mu3e:2020gyw}; more details on our modelling of these effects are described in our companion paper \cite{Mu3e-angles}. 

We find that the invariant masses of the two $e^+e^-$ combinations are peaked towards low values, as shown in Fig.~\ref{fig:mee}. This is because the amplitude represented by the diagram in Fig.~\ref{fig:diagram} is strongly enhanced for an intermediate photon with low virtuality. This behaviour is present in both signal and background, after selecting only low $m_{\slashed{E}}$ events for the latter. This indicates that all three tracks tend to be highly collimated, recoiling against the missing momentum. Thankfully, the strong $B$-field of \muthreee and the low $p_T$ of the tracks mean that even very collimated tracks are being split significantly by the time they reach the tracking layers. Following, \cite{Perrevoort:2018okj} we reject events for which both positrons are within 0.14 radian from each other and at the same time have a momenta which differ with less than \SI{1}{\MeV}. In practice, we find this loss in acceptance to be negligible ($<1\%$) for both signal and background.

\bibliographystyle{JHEP}
\bibliography{bibliography.bib}

\end{document}